# Observation of backscattering-immune chiral electromagnetic modes without time reversal breaking

Wen-Jie Chen[1,*], Jian-Wen Dong[1,2 *,†], Zhi Hong Hang[2], Xiao Xiao[2],

He-Zhou Wang[1,§], and C. T. Chan[2]

[1.] State Key Laboratory of Optoelectronic Materials and Technologies, Sun Yat-Sen (Zhongshan) University, Guangzhou 510275, China

[2.] Department of Physics, The Hong Kong University of Science and Technology, and the William Mong Institute of Nano Science & Technology, Hong Kong, China

* Equal contributions

† Email: dongjwen@mail.sysu.edu.cn

§ Email: stswhz@mail.sysu.edu.cn

A strategy is proposed to realize robust transport in time reversal invariant photonic system. Using numerical simulation and microwave experiment, we demonstrate that a chiral guided mode in the channel of a three-dimensional dielectric woodpile photonic crystal is immune to the scattering of a square patch of metal or dielectric inserted to block the channel. The chirality based robust transport can be realized in nonmagnetic dielectric materials without any external field.

Robust transport behavior of chiral edge states in quantum Hall effect systems is one of the most intriguing phenomena in condensed matter physics [1]. Recently, chiral edge states in photonic systems are theoretically predicted [2] and experimentally realized [3] in two-dimensional magneto-optical photonic crystals (PCs) with strong external magnetic field. The breaking of time reversal symmetry (TRS) is crucial in achieving these one-way transports.

Robust transport behavior can also be found in electronic systems without TRS breaking. Examples are the spin-filtered edge states of the quantum spin Hall effect [4] and surface states of three-dimensional (3D) topological insulators [5]. The robustness of these states is topologically protected. But as Kramers' degeneracy and spin-orbit coupling are specific to electronic systems, it is not obvious that we can realize scattering-immune states in photonic systems without TRS breaking. The purpose of this paper is to show that robust transport in TRS invariant photonic system can indeed be observed simply in a channel drilled inside a layer-by-layer "woodpile" photonic crystal, which is the most common configuration to realize a complete photonic band gap in 3D [6]. Microwave experiments and numerical simulations demonstrate that high transmittance will occur in such a channel even if the channel is blocked by a square slab of perfect electric conductor (PEC) obstacle with size larger than the channel's cross-section. The electromagnetic (EM) wave will go around the PEC obstacle and keep moving forward.

Consider a general non-chiral (NC) 3D PC made with lossless dielectrics and the PC possesses both inversion and mirror symmetries. The bulk modes are linearly polarized and the PC should not differentiate between right-handed (RH) and left-handed (LH) circularly polarized incident wave. By RH or LH, we mean that electric field vectors of the EM wave form an RH or LH helix along the propagation direction at a fixed instant of time. Now suppose that (we will show "how" later) we can create inside this NC PC a waveguide channel that possesses neither inversion nor mirror symmetry, the RH/LH "handed" degeneracy of waveguide modes can be lifted, and thus the waveguide mode will have chirality. For a pair of counter-propagating chiral modes with the same handedness, their coupling is suppressed as the temporal rotation of the electric field at a given point in space has opposite directions. Besides, if such chiral mode is the only allowed mode in a certain frequency region, the channel effectively possesses a polarization gap (PG) that allows only one kind of circularly polarized wave to propagate. When a circularly polarized wave encounters a metal or dielectric slab, the reflected backward wave will have the same temporal rotational direction as the incident wave since no relative phase change occurs in the electric field vector upon reflection. Therefore, the reflected circularly polarized wave will have the opposite handedness which is forbidden in the channel [7]. The EM waves have no choice but to move forward by going around the obstacle resulting in backscattering-immune transport.

Now we will construct such a chiral channel inside a NC 3D woodpile PC which is built by stacking bars with square cross-sections $b \times b$ (yellow) as illustrated in Fig. 1(a). The periodicity along the z direction is $a$, and on the x-y plane is $d$. While the woodpile PC is NC, it can also be viewed as an assembly of discrete chiral ladders twisting upwards in the z-direction as illustrated by the blue-colored building blocks in Fig. 1(a). From this point of view, four chiral ladders with their axes along the z direction can be identified in the woodpile PC, as shown in Fig. 1(b). The handedness of each chiral ladder is illustrated by colored square arrows in the bottom plane of Fig. 1(b). We can see that there are two LH (blue and cyan) and two RH (red and pink) chiral ladders, and the phase difference between two nearest ladder of same-handedness helices is $\pi$. When we drill a void channel with a square cross-section along the z direction, the channel can either be NC or chiral, depending on where we position the channel's central axis. For example, the axis of an NC channel can pass through the star in the bottom plane of Fig. 1(c). Note that it is equivalent to the crossing point between two neighboring perpendicular bars [7]. So an NC channel can be constructed by removing bar segments in two neighboring layers of a period along the z direction (see e.g. the missing part in the scheme of Fig. 1(c)). On the other hand, we can choose to create a channel by removing one chiral ladder, such as the blue LH illustrated in Fig. 1(d). The channel is obviously chiral with its central axis passing through the star in the bottom plane of Fig. 1(d). Note that the NC channel has two mirror symmetric planes with respect to the x and y axes, whereas the chiral channel has neither mirror nor inversion symmetry, but has four-fold helical symmetry.

In the experiment, the sample is constructed by stacking square alumina bars ($\varepsilon = 9$) layer by layer. The sample is surrounded by plexiglass for support. A photograph is shown in Fig. 1(e). Here, $a = 16mm, b = 4mm, d = 11.32mm$, and the thickness of the sample is 160 mm. Photographs of NC and chiral channels are shown in the top and bottom inset of Fig. 1(e) respectively. Figure 1(f) shows the illustration of the experimental setup. The microwave is emitted through an X11644A waveguide from 8.2 to 12.4 GHz with a polarization along the x direction. The source is placed 16 mm below the entrance of the channel. A 15-mm-long dipole antenna is placed at the exit of the channel to measure the transmitted amplitude and phase. By aligning the dipole antenna along either x or y direction, the transmitted amplitude of either x or y polarization as well as the relative phase between the x and y polarization can be measured.

Figure 2(a) shows the projected band structure along the z direction for the structure with the NC channel, calculated by plane wave expansion method [8]. We observe that there is a complete gap in the projected band structure. Three guided modes can be found in the gap, labeled as NC-1, NC-2, and NC-3, with NC-2 and NC-3 being degenerate. Figure 2(b) shows the good agreement between experiments (green) and finite difference time domain (FDTD) simulation (black) [9] on the total transmission spectra of the finite sized sample. The high transmission in the frequency range from 9.8 to 10.3 GHz is consistent with the existence of guided modes in the band gap of the projected bands. These guided modes do not exhibit any chirality due to mirror symmetry of the NC channel. Figure 2(c) shows that measured transmitted amplitudes with the electric field polarized along the x direction ($S_x$) is much larger than that along the y direction ($S_y$). We note that there is still some low transmittance in $S_y$ due to device noise and the fact that the channel is not right at the center of the NC sample, leading to a small symmetry breaking.

We now consider a chiral channel. Figure 3(a) shows the projected band structure along the z direction showing four guided modes in the complete gap. Eigenmode analysis reveals that two of them are LH polarized and two are RH polarized, as indicated by different symbols in Fig. 3(a). Figure 3(b) shows the total transmission spectra of the channel. Both FDTD simulation (black) and measured transmission (green) are consistent with the band structure. Most importantly, we find that there is a PG (light blue region), from 9.76 to 10.12 GHz, in which only LH-polarized guided mode can propagate. The propagation of RH-polarized wave should be attenuated within this PG, allowing for the possibility of robust transport in this frequency region. Inside the PG, the transmitted wave should be LH polarized, even if the incidence wave is linearly polarized. This is verified by the experimental results. Figure 3(c) shows that the spectra of the transmitted amplitudes $S_x$ and $S_y$ are almost the same, and Fig. 3(d) shows that the relative phase are nearly 90° in the PG (light blue region). The relative phase deviates from 90° at the PG edge due to the appearance of the elliptical polarization [7]. Eigenmode analysis shows that the transmitted waves are elliptically polarized outside the PG from 9.34 to 9.76 GHz, where more than one guided modes co-exist [7]. This is also verified by experiment results. We find that the measured transmitted amplitudes $S_x$ and $S_y$ are no longer the same (Fig. 3(c)), while the relative phases with the value of around 315° have larger fluctuations than those in the PG (Fig. 3(d)).

Now, we come to demonstrate the backscattering-immune property of the chiral channel states. We note that only LH guided modes are allowed to propagate in chiral channel within the PG frequency region. When an LH polarized waves move forward and encounter a square PEC obstacle as illustrated in Fig. 4(b), the reflection will change LH to RH polarization. The reflected wave cannot propagate due to the lack of the backward guided mode. Poynting vector patterns of such a backscattering-immune state from FDTD simulations are shown in Fig. 4. The simulated geometry parameters are the same as those in the experiment and an $E_x$-polarized source at 9.9 GHz is used. In order to see clearly inside the channel, we just show the patterns within a period of $2a$. Figures 4(a) and 4(b) illustrate the results without and with a PEC slab (with the size of $12mm \times 12mm$, gold color in Fig. 4(b)), respectively. It is found that flux propagates in the unblocked channel from bottom to top in a helical shape inside the LH chiral channel. When the slab is inserted into the chiral channel, the flux circumvents the PEC boundary and continues moving upward, showing the robust transport phenomenon and thus high transmittance.

For completeness, we also did FDTD simulations for the NC channel with the experimental geometry. The obstacle is the same as that in Fig. 4(b) and the working frequency is 10.2 GHz. As the NC channel has mirror symmetry, the allowed channel modes do not exhibit any chirality. FDTD results in Fig. 4(c) show that linear polarized EM wave propagates from bottom to top in the channel and the energy flow is blocked once the PEC slab is inserted in the channel, as shown in Fig. 4(d).

In order to verify the simulations, we have measured the transmissions with different obstacles inserted inside the channels. Figure 5(a) shows the results of the chiral channel for the frequency region where the robust transport is present. The first obstacle is an aluminum slab with a size of $6.6mm \times 6.6mm \times 2.5mm$, approximately the same as the channel's cross-section. We can see that forward transmission (red) is almost the same as that of the unblocked channel (black). If a larger PEC block with nearly double the size is inserted, the transmission (green) is only slightly decreased. These observations verified the robust transport character of the chiral state. Such robust transport cannot be observed outside the PG for the metal obstacle. Low transmission is recorded outside the PG since the EM wave is strongly reflected by the metal blocks. For the case of a dielectric slab, the robust transport effect can also be observed inside the PG. When an alumina slab with the same size as the first PEC slab is inserted, the feature of robust transport is maintained as the transmissions remain high (blue dash). Note that the spectrum is slightly different from that of the unblocked channel due to the Fabry-Perot interference effect. The FDTD simulation shows that the EM waves will penetrate through the dielectric slab instead of going around it [7]. For the purpose of comparison, we also measured the transmission spectra in the blocked/unblocked NC channel. Low transmissions were observed in the spectra for either the dielectric or PEC blocks (Fig. 5(b)). We found that the EM wave was strongly reflected by the blocks, which agreed well with our simulations.

Many NC PCs contain chiral building blocks [10] and the guided modes can have "handedness" if channels are drilled in these PCs. The necessary and sufficient conditions for the backscattering-immune tunneling state are the fact that the photonic structure should have guided modes of only one chirality in 3D photonic bandgaps and the obstacle reflects waves into the opposite chirality.

In conclusion, we realized robust transport in TRS invariant photonic crystals by creating a chiral channel in a 3D dielectric PC. Both numerical simulation and microwave experiment demonstrated that a chiral channel mode is robust against scattering of obstacles that reflect RH waves into LH waves. It can be viewed as phenomenologically similar to the topologically protected modes in TRS invariant topological insulators. The fact that the robust transport can be realized in a dielectric PC and does not require an external field is an advantage over two-dimensional PC configurations which requires magnetic materials and external magnetic field to break TRS.

We thank Prof. Z. F. Lin and Dr. J. C. W. Lee for their thoughtful input. Work in Guangzhou was supported by NSFC grants (10804131, 11074311, 10874250), FRFCU grant (2009300003161450), and GDNSF grant (10451027501005073). Work in Hong Kong was supported by Hong Kong RGC grant (600209), and computation resources were supported by the Shun Hing Education and Charity Fund. We thank Prof. W. J. Wen for the microwave equipments.

**References**

[1] See e.g. K. von Klitzing, et al., Phys. Rev. Lett. **45**, 494 (1980); R. E. Prange and S. M. Girvin, The Quantum Hall Effect (Springer, New York, 1987); Y. Zhang, et al., Nature **438**, 201 (2005).

[2] F. D. M. Haldane and S. Raghu, Phys. Rev. Lett. **100**, 013904 (2008); Z. Wang, et al., Phys. Rev. Lett. **100**, 013905 (2008); X. Y. Ao, et al., Phys. Rev. B **80**, 033105 (2009).

[3] Z. Wang, et al., Nature **461**, 772 (2009); J. X. Fu, et al., Appl. Phys. Lett. **97**, 041112 (2010); Y. Poo, et al., Phys. Rev. Lett. **106**, 093903 (2011).

[4] C. L. Kane and E. J. Mele, Phys. Rev. Lett. **95**, 226801 (2005).

[5] See e.g. L. Fu, et al., Phys. Rev. Lett. **98**, 106803 (2007); D. Hsieh, et al., Nature **452**, 970 (2008); Y. L. Chen, et al., Science **325**, 178 (2009); M. Z. Hasan and C. L. Kane, Rev. Mod. Phys. **82**, 3045 (2010) and reference therein.

[6] K. M. Ho, et al., Solid State Commun. **89**, 413 (1994). E. Özbay, et al., Phys. Rev.

B **50**, 1945 (1994).

[7] See Supporting Materials.

[8] S. G. Johnson and J. D. Joannopoulos, Opt. Express **8**, 173 (2001).

[9] A. F. Oskooi, et al., Comput. Phys. Commun. **181**, 687 (2010).

[10] A. Chutinan and S. Noda, Phys. Rev. B **57**, R2006 (1998).

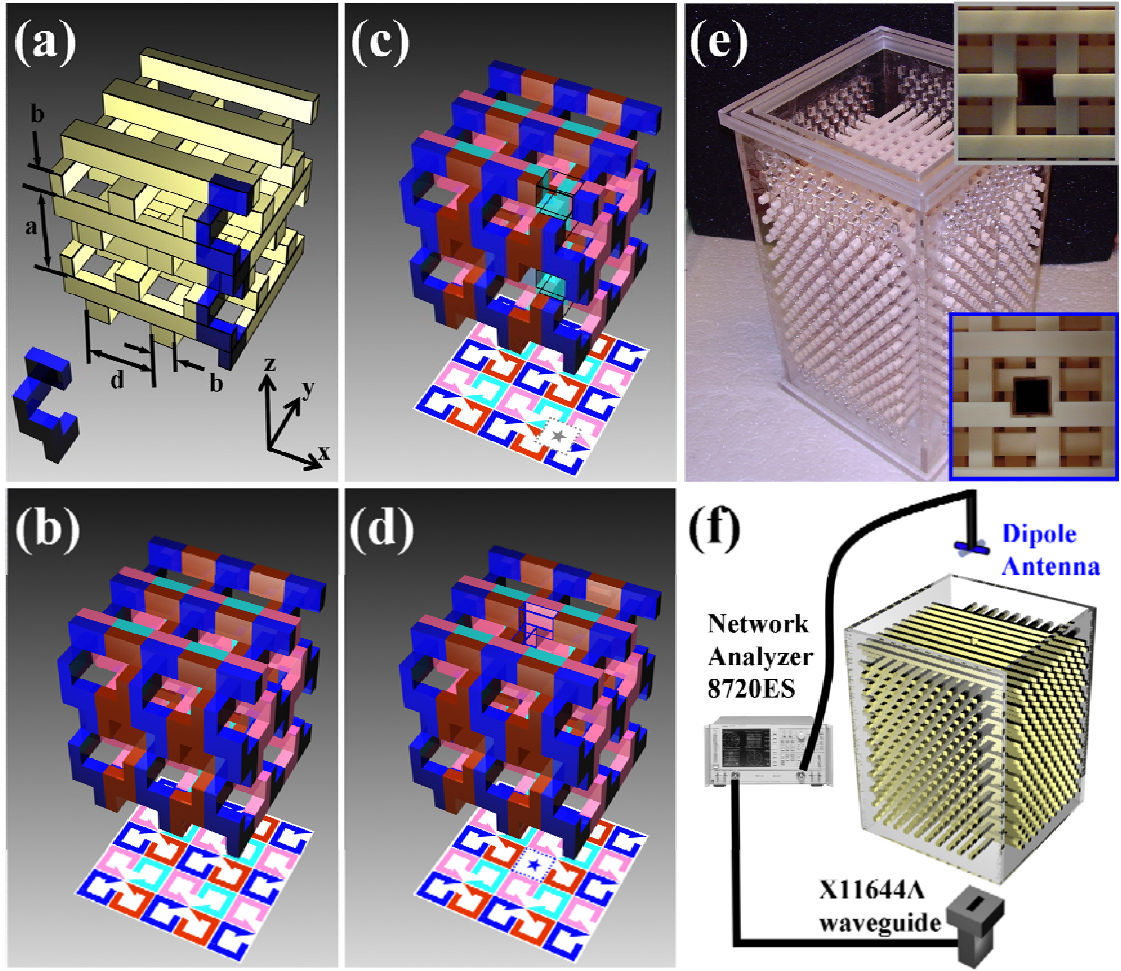

FIG. 1. (Color online) (a) PC constructed by square bars. One chiral ladder is marked in blue to guide the eye. (b) Woodpile PC comprising two LH ladders (blue and cyan) and two RH ladders (red and pink). (c)/(d) NC/chiral channel constructed by removing material along the z direction. Colored arrows in the bottom plane of (b)-(d) show the handedness of the ladders. Stars in (c) and (d) show the position of the channel's central axis. (e) Photograph of the experimental sample. Insets show the exits of the NC (top) and chiral (bottom) channel. (f) Setup for transport properties measurement. Surrounding transparent plexiglass walls are used for supporting the sample.

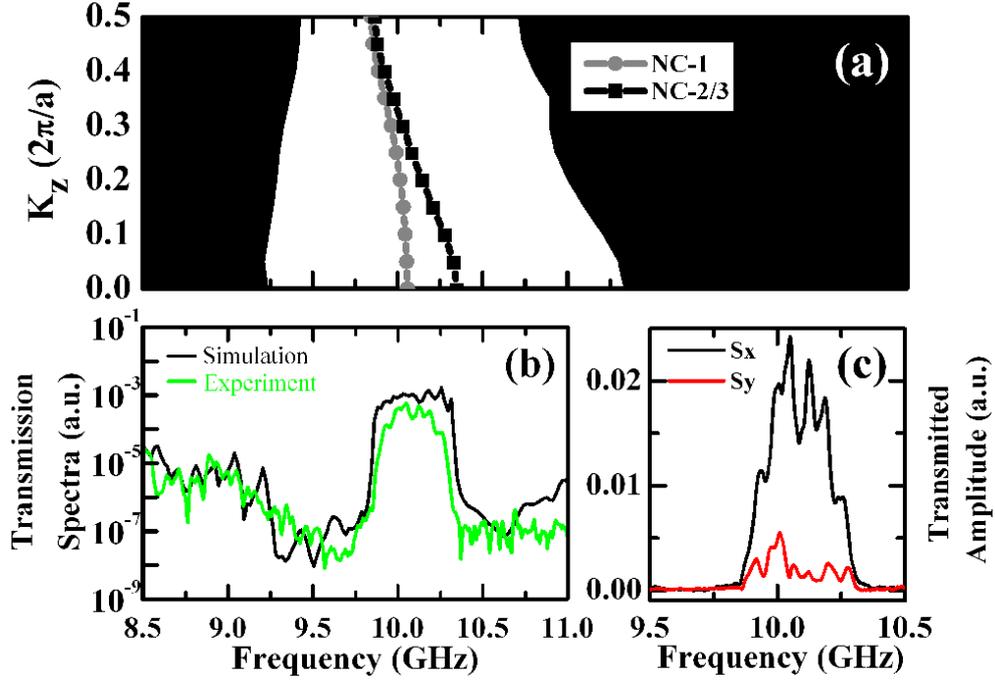

FIG. 2. (Color online) Propagation characteristics for the NC channel. (a) Band structure (solid symbols: NC guided modes; black: projected bulk bands). (b) Total transmission spectra (black: simulation, green: experiment). (c) Measured transmitted amplitudes of $S_x$/$S_y$ (black/red).

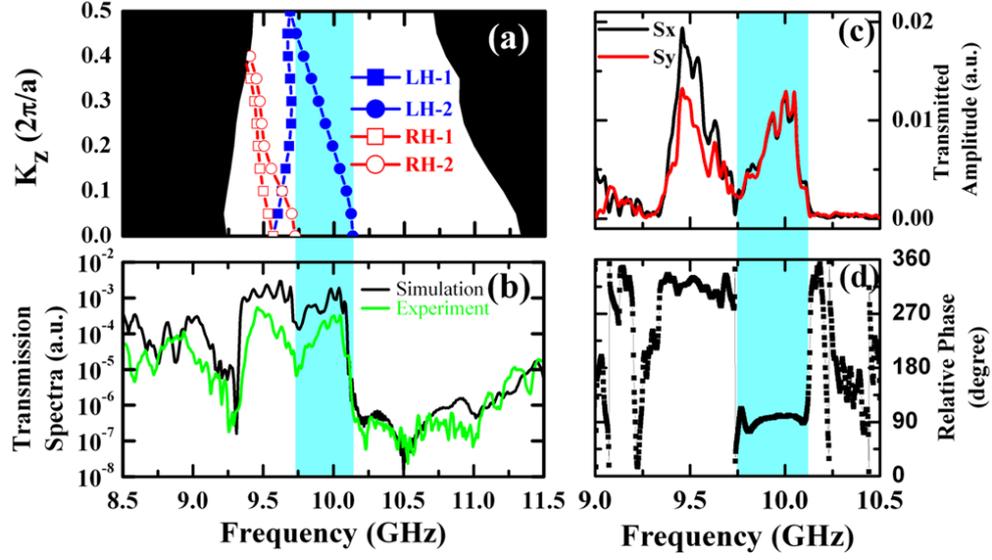

FIG. 3. (Color online) Propagation characteristics for the chiral channel. (a) Band structure (blue solid: LH guided modes, red open: RH guided modes, black: projected pass bands). (b) Total transmission spectra (black: simulation, green: experiment). (c) Measured transmitted amplitudes of $S_x/S_y$ (black/red). (d) Relative phase between $S_x$ and $S_y$. Light blue highlights the region for the poloarization gap.

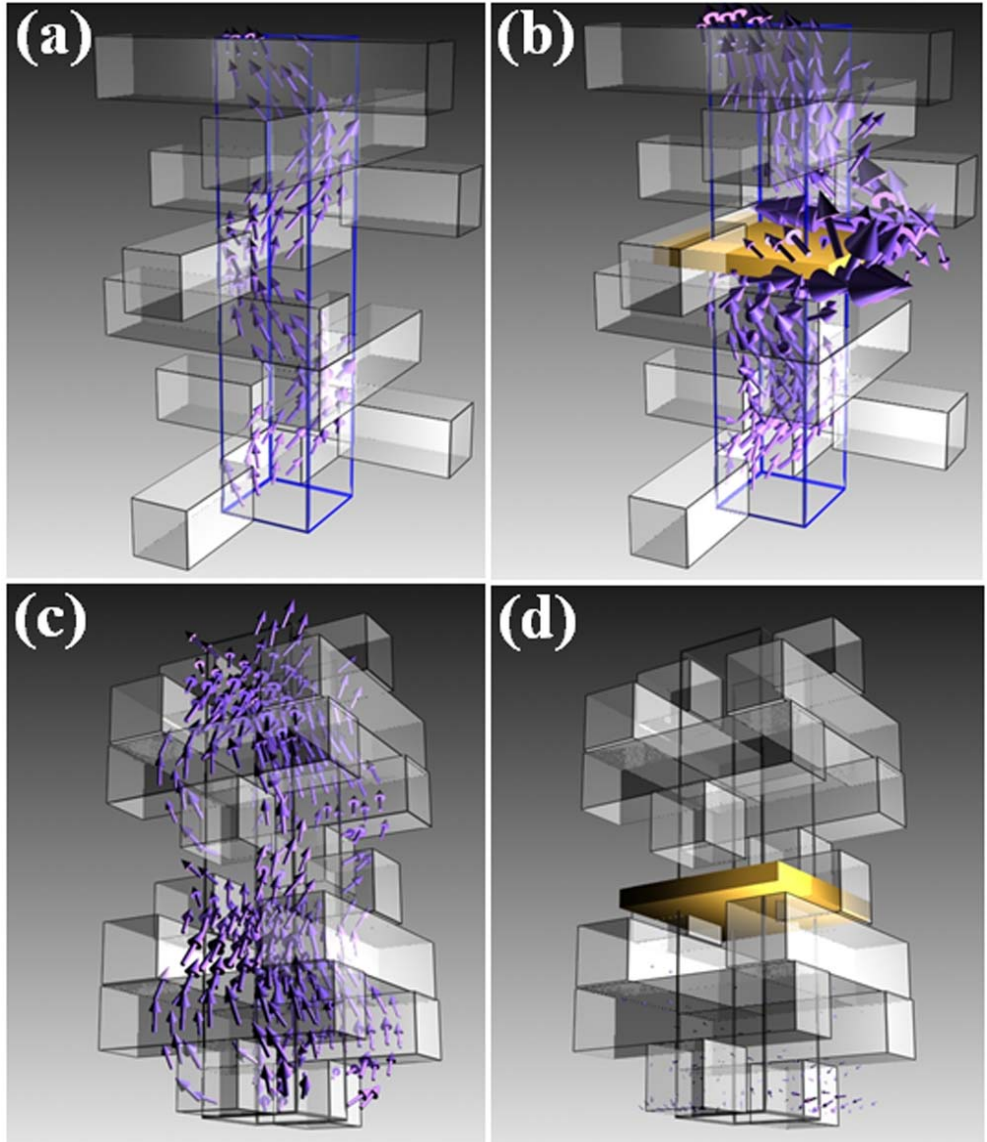

FIG. 4. (Color online) Poynting vector patterns (violet arrows) for (a)/(b) the unblocked/PEC-blocked chiral channel, and (c)/(d) unblocked/PEC-blocked NC channel. Gold color represents the PEC plate. Blue/gray frames guide eyes to see the boundary of waveguide channel. Transparent bars represent the geometrics near the channel.

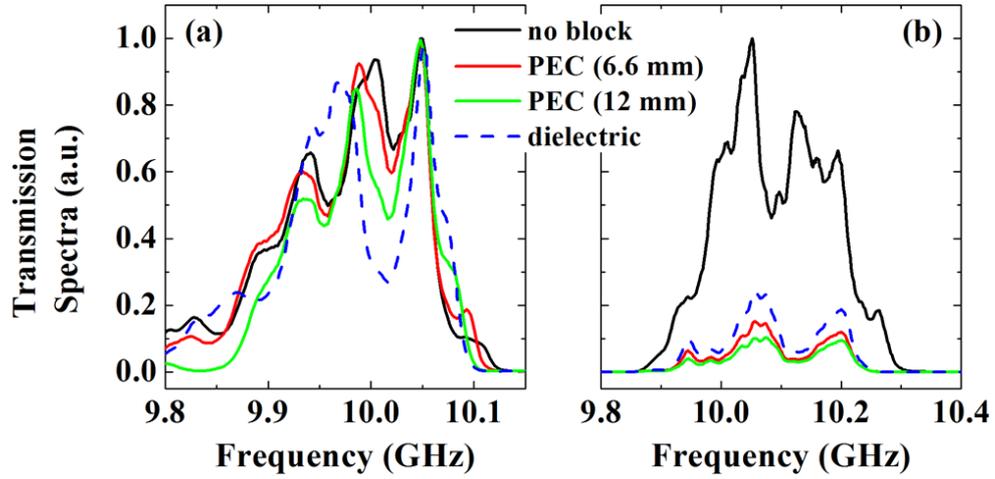

FIG. 5. (Color online) Measured total transmittance for the (a) presence and (b) absence of backscattering-immune transport. Black solid line: no obstacle in channel. Red/green solid line: small/large PEC plate in channel. Blue dash line: dielectric plate in channel.